# Violation of the sphericity assumption and its effect on Type-I error rates in repeated measures ANOVA and multi-level linear models (MLM)

**Nicolas Haverkamp**[1, a], **André Beauducel**[1, b]

[1]University of Bonn, Institute of Psychology, Kaiser-Karl-Ring 9, 53111 Bonn, Germany

[a]nicolas.haverkamp@uni-bonn.de, [b]beauducel@uni-bonn.de

April 25[th], 2017

**Abstract.** This study aims to investigate the effects of violations of the sphericity assumption on Type I error rates for different methodical approaches of repeated measures analysis using a simulation approach. In contrast to previous simulation studies on this topic, up to nine measurement occasions were considered.

Therefore, two populations representing the conditions of a violation vs. a non-violation of the sphericity assumption without any between-group effect or within-subject effect were created and 5,000 random samples of each population were drawn. Finally, the mean Type I error rates for Multilevel linear models (MLM) with an unstructured covariance matrix (MLM-UN), MLM with compound-symmetry (MLM-CS) and for repeated measures analysis of variance (rANOVA) models (without correction, with Greenhouse-Geisser-correction, and Huynh-Feldt-correction) were computed. To examine the effect of both the sample size and the number of measurement occasions, sample sizes of $n = 20, 40, 60, 80,$ and $100$ were considered as well as measurement occasions of $m = 3, 6$ and $9$.

For MLM-UN, the results illustrate a massive progressive bias for small sample sizes ($n = 20$) and $m = 6$ or more measurement occasions. This effect could not be found in previous simulation studies with a smaller number of measurement occasions. The mean Type I error rates for rANOVA with Greenhouse-Geisser-correction demonstrate a small conservative bias if sphericity was not violated, sample sizes were small ($n = 20$), and $m = 6$ or more measurement occasions were conducted.

The results plead for a use of rANOVA with Huynh-Feldt-correction, especially when the sphericity assumption is violated, the sample size is rather small and the number of measurement occasions is large. MLM-UN may be used when the sphericity assumption is violated and when sample sizes are large.

**Keywords:** Multilevel linear models, mixed linear models, hierarchical linear models, repeated measures ANOVA, simulation study

**Introduction**

Multilevel linear models (MLM) have been discussed as an alternative to repeated measures analysis of variance (rANOVA; Arnau et al., 2010; Goedert, Boston, Barrett, 2013; Gueorguieva & Krystal, 2004) and, sometimes, researchers have even been urged to use MLM instead of rANOVA (Boisgontier & Cheval, 2016). Although MLM are increasingly used instead of rANOVA, the terminology is heterogeneous in this area. Different labels are used to denote models encompassing fixed and random effects, covariance pattern models, and regression models that are based on more than one data level, where the levels are typically defined by the measurement occasions nested within individuals. These models are referred to as hierarchical linear models (Bryk & Raudenbush, 1992; Raudenbush & Bryk, 2002), as (general) mixed linear models (Arnau, Balluerka, Bono, Gorostiaga, 2010; McLean, Sanders, & Stroup, 1991), mixed effects models (Gueorguieva & Krystal, 2004), multilevel linear models (Hox, 2002), or as multilevel models (Lucas, 2014; Maas & Hox, 2005). We will follow Tabachnick and Fidell (2013) in using MLM to denote these models in the following.

Here, we highlight three advantages of MLM over rANOVA: First, MLM allows to model data that correspond to a multi-level structure. Whenever researchers assume at least one level of data being nested within another level, MLM is appropriate. If there is only one level of measurement occasions and one level of individuals, rANOVA is also an appropriate model, but for any more complex structure comprising several levels MLM will be more appropriate than rANOVA (Baayen, Davidson & Bates, 2008). Second, MLM is robust even when there are several randomly distributed missing values. Especially in research designs with several measurement occasions a large number of missing values can occur. Since parameters (e.g., slope parameters) are estimated for each individual, there is no requirement for complete data over occasions. A third advantage is the possibility to compare MLM with different assumptions on the covariance structure of the data. For example, models with auto-regressive covariance structure, with uncorrelated structure, or with compound-symmetry (CS) are possible. When there are no specific assumptions on the covariances, MLM with an unstructured covariance (UN) matrix can be specified. To summarize, MLM has substantial advantages when compared to rANOVA.

Given the important advantages of MLM over rANOVA, MLM will be the method of choice for several repeated measures designs. Of course, the appropriateness of MLM results also depends on data characteristics, modeling options, and estimation procedures (Lucas, 2014; Tabachnick & Fidell, 2013), so that simulation studies are necessary to ascertain the quality of MLM results. Since several aspects can be varied in MLM, a single simulation study cannot cover all relevant aspects, so that different studies with a focus on different aspects have been performed. With respect to the estimation method (e.g., maximum likelihood versus restricted maximum likelihood), some simulation studies indicate that restricted maximum likelihood is more accurate (Maas & Hox, 2005), but it seems that restricted maximum likelihood is superior for random effects and not necessarily for fixed effects (West, Welch, & Galecki, 2007). Further results and recommendations are available with regard to sample size, group size, and number of groups. However, Maas and Hox (2005) noted that some inconsistencies of their results on sample size with the results of other simulation studies were probably related to the use of different simulation designs and different simulated conditions. The multiple options that are possible with MLM, the flexibility of the method, may have enhanced the specificity of results. For example, the statistical power and Type I error rate of MLM based on autoregressive covariances and based on unstructured covariances has been investigated by Gueorguieva and Krystal (2004). They found the lowest Type I error rate for the MLM based on autoregressive covariances. It should, however, be noted that their simulated data had an autoregressive structure. This indicates that an optimal Type I error control results when the covariance structure specified in MLM corresponds to the empirical covariance structure. In contrast, Kowalchuk, Keselman, and Algina (2004) found that MLM based on UN performed similarly to fitting the true covariance structure and, under certain conditions, showed better Type I error control. Thus, the results of simulation studies on the specification of MLM for obtaining optimal Type I error rates are not conclusive.

Goedert et al. (2013) performed a simulation study with another type of data containing violations of the sphericity assumption and found a superior statistical power of MLM with unstructured covariances (UN) when compared to rANOVA. In their simulation study, the MLM based on the $F$-distribution with between-within degrees-of-freedom (West et al., 2007) showed no bias in Type I error rates, whereas MLM-UN based on Wald's $z$ had led to large Type I error rates. On this basis, Goedert et al. (2013) recommended the use of MLM-UN based on the $F$-statistic instead of MLM-UN based on $z$ or rANOVA, especially for small samples. We followed their recommendation in performing MLM-UN based on $F$ in the following. In Goedert et al. (2013), the Type I error rates of MLM-UN based on $F$ and rANOVA with Greenhouse-Geisser (GG) correction (Greenhouse & Geisser, 1959) were rather similar with six measurement occasions and a sample size of at least 30 cases.

Gueorguieva and Krystal (2004) also found widely acceptable and similar Type I error rates for MLM-UN and rANOVA-GG in a simulation study based on four measurement occasions. However, in their simulation study the Type I error rates were slightly more correct for MLM based on compound symmetry (CS). It should be noted that the CS assumption is related to the sphericity assumption of rANOVA. The CS assumption is more restrictive than the sphericity assumption (Field, 1998) so that MLM with CS will also satisfy the sphericity assumption. It would therefore be of interest to compare MLM based on CS with rANOVA results. Since rANOVA is always based on the sphericity assumption, it will also be important to compare MLM based on CS, which is more restrictive than rANOVA, with MLM based on UN, which is less restrictive than rANOVA. In order to provide a comprehensive description of the effects, a comparison of MLM and rANOVA should be based on data sets that satisfy the CS assumption as well as the sphericity assumption and it should furthermore be based on data-sets that violate both the CS assumption and the sphericity assumption. Data that are conform to the sphericity assumption and that simultaneously violate the CS assumption are very specific (Field, 1998) and will therefore occur very rarely so that they are not interesting for a comparison of MLM with rANOVA. Since violations of the sphericity assumption are known to result in progressive Type I error rates of rANOVA, the rANOVA-GG has been proposed in order to compensate for the progressive bias. The probably less conservative correction of the progressive bias of Type I error rates of rANOVA proposed by Huynh and Feldt (HF; 1976) should also be considered.

Since the focus of Goedert et al. (2013) was laid on data relevant for research on spatial neglect, they only investigated data in which the sphericity assumption and the CS assumption were violated. The aim of the present simulation study was to extend their results on Type I error rates beyond the specific data characteristics that are relevant for research on spatial neglect. Gueorguieva and Krystal (2004) also simulated data with a mild violation of the sphericity assumption data because they investigated an autoregressive covariance structure. We will therefore investigate whether the results provided by Goedert et al. (2013) as well as Gueorguieva and Krystal (2004) can also be found for data that are conform to the sphericity assumption. Accordingly, we will compare MLM and rANOVA for data sets with and without violation of the sphericity assumption and for MLM based on UN and CS.

Simulation studies that are based on rather specific models, options, and data yield rather specific results, which might, of course, be relevant for the respective field of research when the MLM used in the simulation corresponds to the MLM that is typically used in the respective research. However, these simulation studies might be complemented with simulation studies with a focus on rather simple models and data, which do not depend so much on a large number of specific modelling options and data characteristics. Even when the focus of simulation studies will always be limited, studies that are based on very simple models and data characteristics may provide a baseline for the evaluation of more complex models. For this reason, the present simulation study will only investigate the abovementioned effects of violation versus non-violation of the sphericity assumption on Type I error rates in MLM-UN, MLM-CS, and rANOVA-models (without correction, with GG-correction, and HF-correction) without any between-group effect. Because rANOVA cannot be used for the simultaneous analysis of several data levels, a comparison of MLM and rANOVA does not make sense for such complex data. The current simulation study is therefore only based on a subset of repeated measures data that can be analyzed by means of rANOVA and MLM. Since the data and models analyzed in this study do not contain fixed between group effects, restricted maximum likelihood estimation will be used, since this estimation method has been shown to be most exact for random effects models (West et al., 2007). The restriction to the class of simple within-subjects models allows for an analysis of up to nine measurement occasions, which has not been done before. Kowalchuk et al. (2004) concluded MLM-UN performs at least as well for Type I error control as MLM with known covariance matrices, which would be MLM-CS for the present data. However, their simulation study was only based on four measurement occasions. Accordingly, a central aim of the present simulation study is to investigate whether the results of Kowalchuk et al. (2004) can be generalized to more than four measurement occasions. Moreover, the abovementioned restrictions allow for 5,000 samples to be drawn from the population in each condition in order to reach substantial robustness of results.

**Material and Methods**

The simulations were performed with IBM SPSS Statistics Version 23.0.0.3. Three factors were varied systematically in the simulation study to investigate their effect on the results of repeated measures analyses: Violations of the sphericity assumption, sample sizes and number of measurement waves. To investigate the 'pure' effect of the respective analysis method on the resulting Type I error rate, neither a between-subject effect nor a within-subject effect were fixed.

Concerning the sphericity assumption, two conditions were established: Under the first condition (sphericity), the sphericity assumption was not violated in the population. A population of normally distributed, z-standardized and uncorrelated variables was generated by means of the SPSS Mersenne Twister random number generator for this condition. Each variable represented the dependent variable measured at one measurement occasion. In the second condition (non-sphericity), the sphericity assumption was violated in the population. For this condition, a population of normally distributed, z-standardized variables was generated. Again, each variable represented the dependent variable measured at one measurement occasion. In order to realize the violation of the sphericity assumption, the correlation between the dependent variables at odd measurement occasions was .80 and the correlation between the dependent variables at even measurement occasions was zero.

As stated before, 5,000 samples were drawn from the populations of generated variables, submitted to all of the different analysis methods (rANOVA without correction, rANOVA with GG-correction, rANOVA with HF-correction, MLM-UN and MLM-CS) and the average Type I error rate was computed for every sample under the two conditions (sphericity vs. non-sphericity) for each of the analysis methods.

Two additional factors that were considered here were the sample size and the number of measurement occasions (waves). Sample sizes were $n = 20, 40, 60, 80,$ and $100$ and measurement occasions were $m = 3, 6$ and $9$. Accordingly, the simulation study comprised 150 conditions (= sphericity[2] × analysis methods[5] × $n$[5] × $m$[3]) with 5,000 samples for each condition. For each condition, the Type I error rate was reported for the .05 alpha-level.

To evaluate whether this simulation is able to deliver stable results, the standard deviation was computed for every average Type I error rate and included in the following graphics which illustrate the simulation findings.

**Results**

The Type I error rates for three measurement occasions and sphericity were close to what was expected at an alpha level of .05 and they were not substantially affected by sample size and method of data analysis (see Figure 1A). For the non-sphericity condition and three measurement waves, the Type I error rates were close to the expectation for all methods except rANOVA and MLM-CS. For rANOVA and MLM-CS, a slight progressive bias was found (see Figure 1B). Note that here and in the following the results for rANOVA and MLM-CS were so similar that the respective lines completely overlap. Again, the effect of sample size was not substantial.

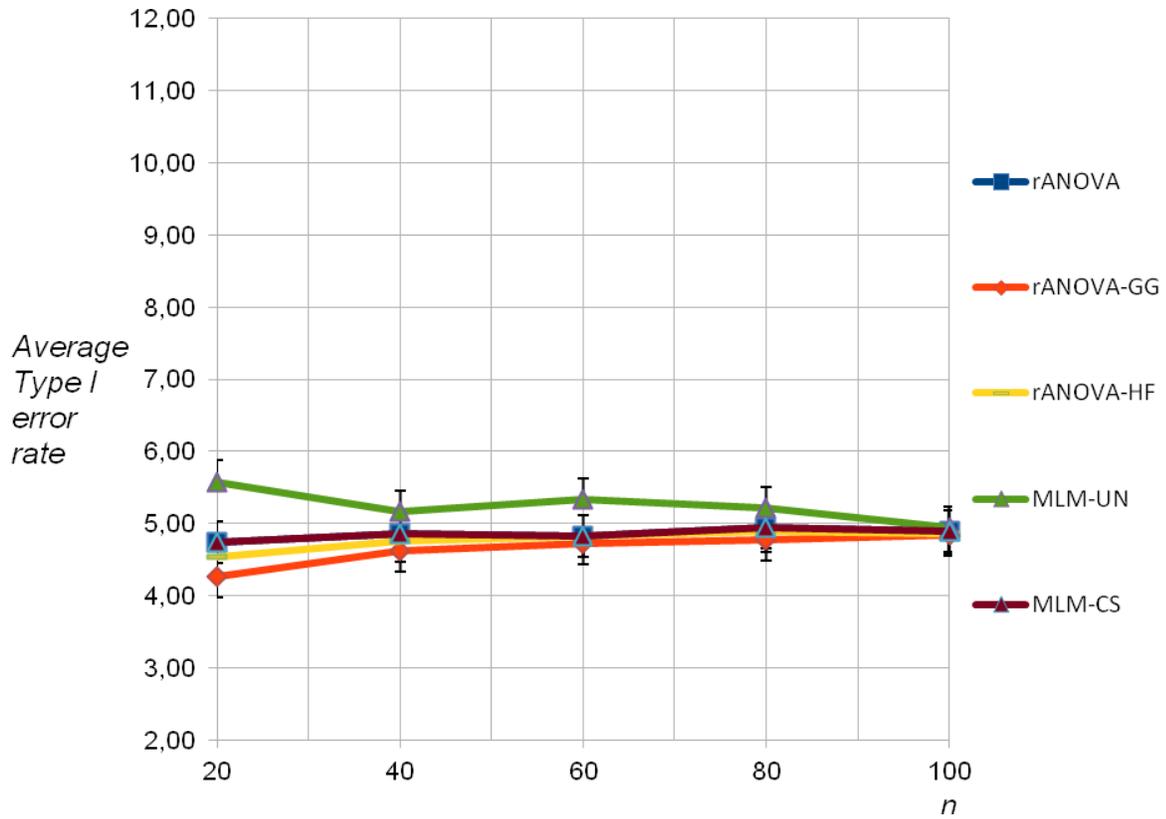

**Figure 1A:** Average Type I error rates for 5,000 tests: no sphericity violation, three measurement waves

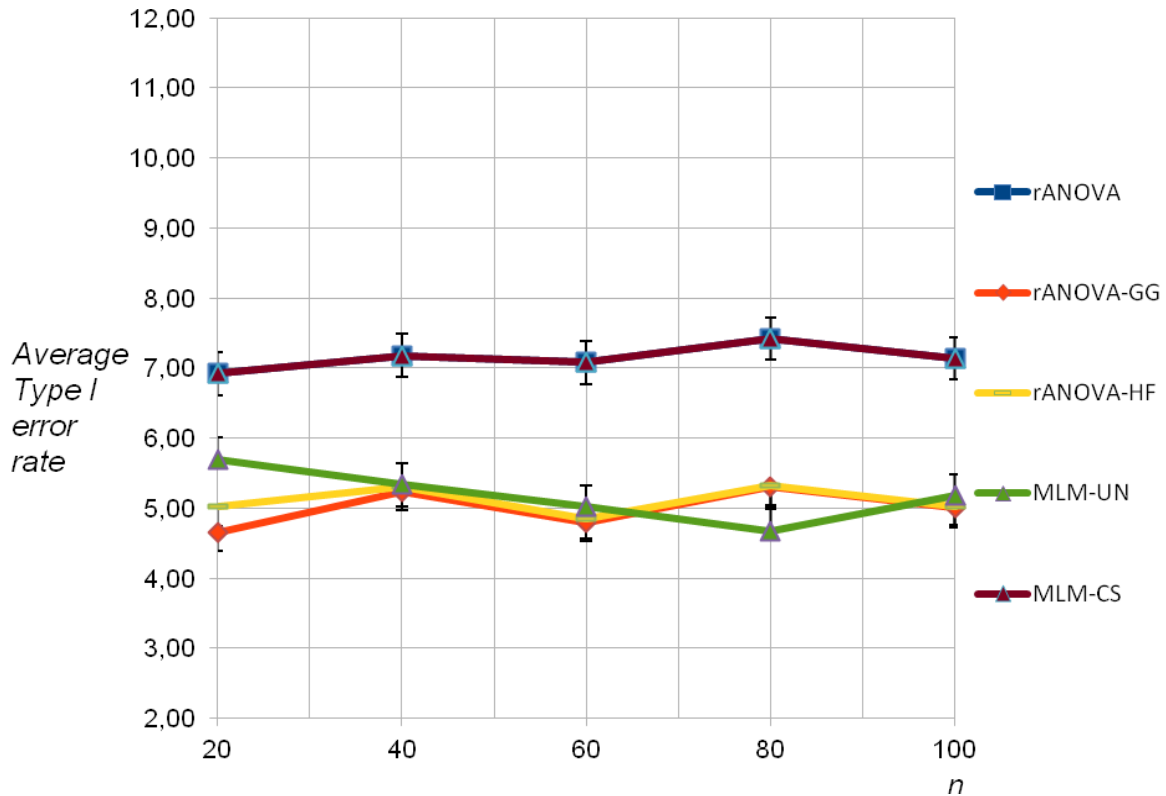

**Figure 1B:** Average Type I error rates for 5,000 tests: sphericity violation, three measurement waves

Type I error rates were as expected for all methods except MLM-UN and rANOVA-GG for the sphericity condition and six measurement occasions. For MLM-UN, a strong progressive bias occurred for $n = 20$, a small progressive bias was found for $n = 40$ and a small conservative bias was found for rANOVA-GG with $n = 20$ (see Figure 2A). In the case of non-sphericity and six measurement waves, there was again a small progressive bias of the Type I error rates with MLM-CS and rANOVA. Moreover, a strong progressive bias occurred for MLM-UN and $n = 20$ and a small progressive bias occurred for MLM-UN with $n = 40$ and $n = 60$ (see Figure 2B).

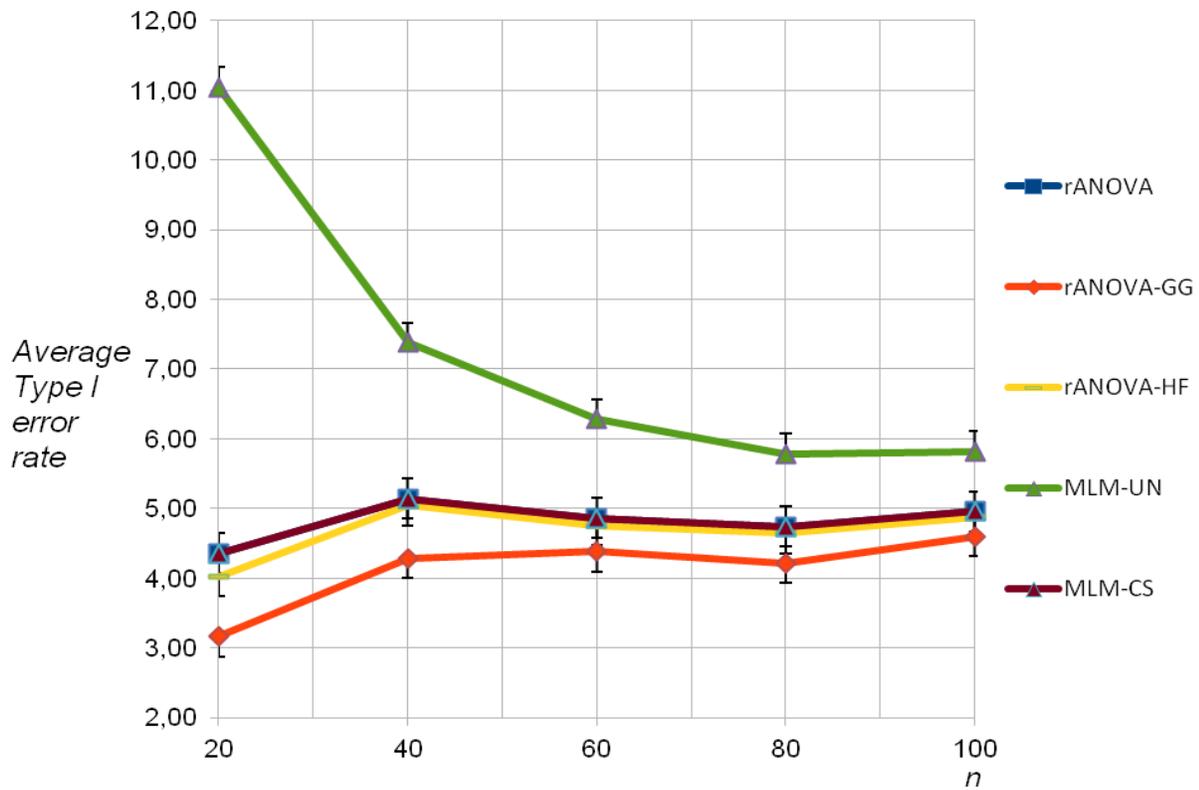

**Figure 2A: Average Type I error rates for 5,000 tests: no sphericity violation, six measurement waves**

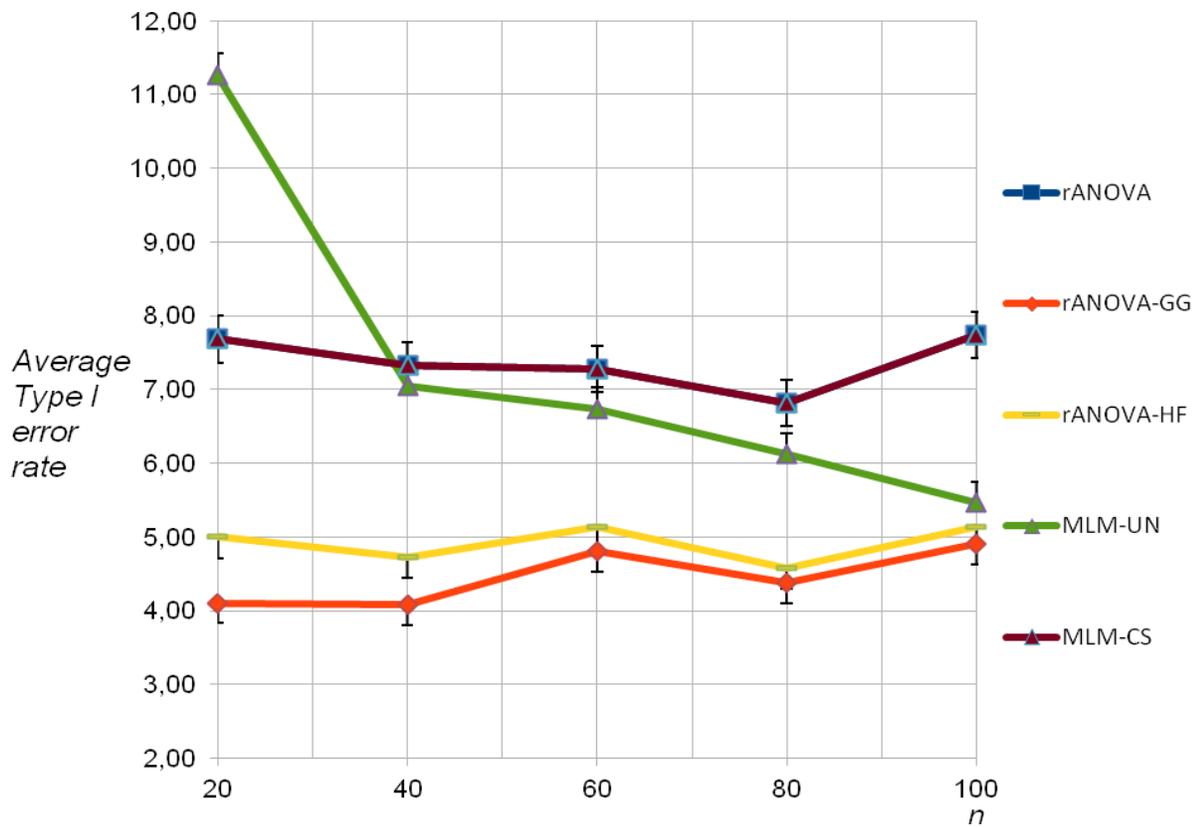

**Figure 2B: Average Type I error rates for 5,000 tests: sphericity violation, six measurement waves**

For the sphericity condition and nine measurement occasions, the mean Type I error rates were again as expected for all methods except MLM-UN and rANOVA-GG: A massive progressive bias was found for MLM-UN with $n = 20$ as well as $n = 40$, while there was still substantial progressive bias for $n = 60$; small conservative biases occurred with rANOVA-GG for $n = 20$ as well as $n = 40$ (see Figure 3A). For the condition of non-sphericity and nine measurement waves, there was a substantial progressive bias of the Type I error rates with MLM-CS and rANOVA. Moreover, a very strong progressive bias occurred for MLM-UN and $n = 20$ as well as for $n = 40$ and a small progressive bias occurred for MLM-UN with $n = 60$ (see Figure 3B).

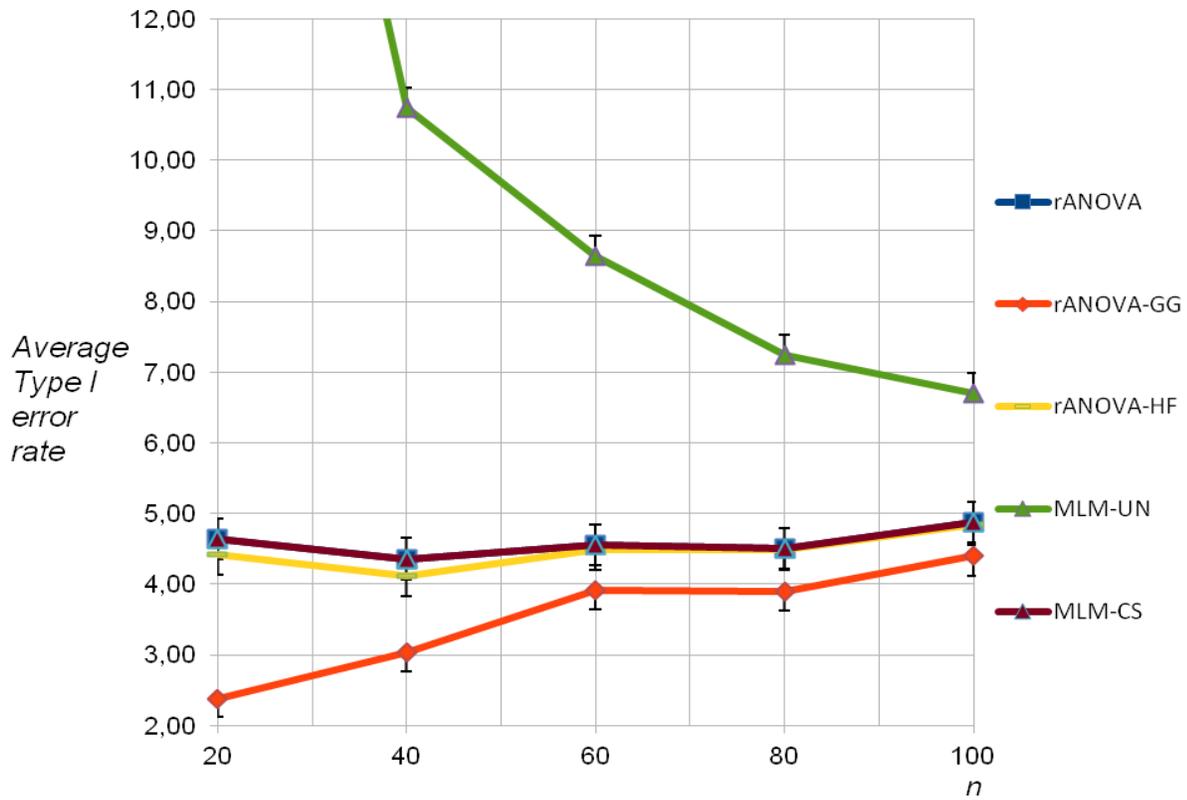

**Figure 3A:** Average Type I error rates for 5,000 tests: no sphericity violation, nine measurement waves (average rate for MLM-UN and n=20: 22.72)

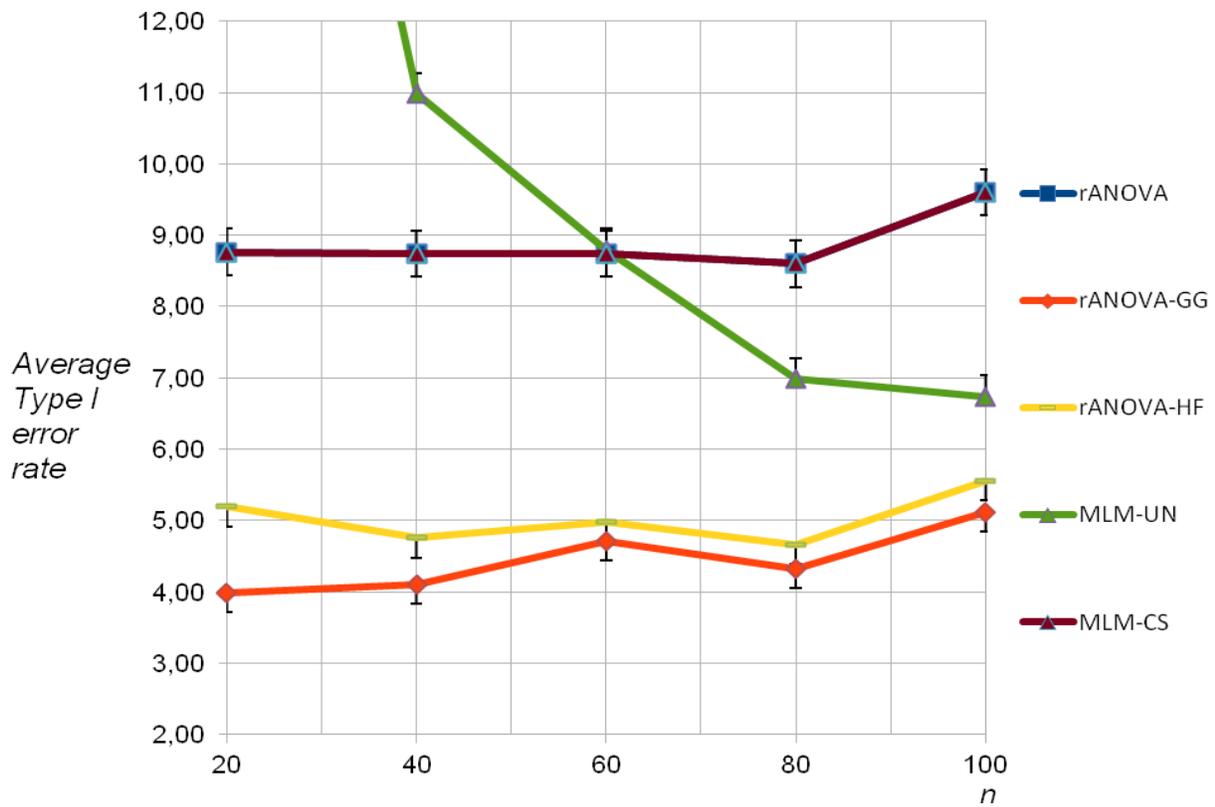

**Figure 3B:** Average Type I error rates for 5,000 tests: sphericity violation, nine measurement waves (average rate for MLM-UN and n=20: 23.18)

**Discussion**

The simulations showed the following results for the mean Type I error rates of the different analysis methods (MLM-UN, MLM-CS, rANOVA without correction, rANOVA-GG and r-ANOVA-HF) under the conditions of violation vs. non-violation of the sphericity assumption and for sample sizes of $n = 20, 40, 60, 80$, and 100 as well as measurement occasions of $m = 3, 6$ and 9: A slight progressive bias for rANOVA and MLM-CS was found in case of a violation of the sphericity assumption. This effect could be demonstrated regardless of the sample size as well as the number of measurement occasions. For MLM-UN, a massive progressive bias for small sample sizes ($n = 20$) and $m = 6$ or more measurement occasions occurred. The progressive bias of MLM-UN was substantial for nine measurement occasions and up to medium sample sizes ($n = 60$). The mean Type I error rates for rANOVA-GG showed a small conservative bias for $m = 6$ or more measurement occasions and small sample sizes ($n = 20$) when the sphericity assumption was not violated.

The most general result of the present simulation study is that there was a substantial progressive bias of Type I error rates for MLM-UN for nine measurement occasions with sample sizes of $n = 60$ and below. This progressive bias for MLM-UN occurred when the population data was conform to the sphericity assumption but it also occurred when the sphericity assumption was violated in the population data. The progressive bias of MLM-UN for nine measurement occasions with sample sizes of $n = 60$ and below was even greater than .075, the upper level of Bradley's (1978) liberal criterion for the evaluation of an empirical estimate $\hat{\alpha}$ of the Type I error rate $(.5\alpha \geq \hat{\alpha} \leq 1.5\alpha)$. It should be noted that MLM was based on the $F$-statistic as recommended by Goedert et al. (2013) because MLM based on Wald's $z$ already showed a progressive bias of Type I error rates in their simulation study. Thus, the present study extends Goedert et al.'s (2013) finding of a progressive bias for MLM to MLM-UN based on a large number of measurement occasions and small sample sizes, even when based on the $F$-statistic. Thus, when there are nine or more measurement occasions and when MLM-UN is used because the sphericity assumption is violated in the data, sample sizes of at least 80 participants should be investigated.

The simulated data of Goedert et al. (2013) and Gueorguieva and Krystal (2004) were based on a violation of the sphericity assumption and on a substantially smaller number of measurement occasions. Therefore, the substantial Type I error rates of MLM-UN in the condition without a violation of the sphericity assumption, nine measurement occasions and sample sizes of $n = 60$ and below could not be found in these simulation studies. Since the Type I error rates of MLM-CS were correct under this condition, this result indicates that MLM-UN should not be used as a form of standard procedure and that a specification of a known covariance structure in MLM might help to avoid progressive bias.

The results for MLM-CS and (uncorrected) rANOVA were so similar across all conditions of the simulation study that there was a total overlap of the respective lines in the figures. Accordingly, when the sphericity assumption was violated in the population data, MLM-CS had the same progressive bias as rANOVA. The progressive bias of MLM-CS and rANOVA increased slightly with the number of measurement occasions so that it was even larger than .075 $(=1.5\alpha)$ for nine measurement occasions. Nevertheless, the progressive bias of MLM-CS and rANOVA was already substantial for three measurement occasions. Accordingly, MLM-CS and rANOVA cannot be recommended when the sphericity assumption is violated. It should be noted that MLM-UN resulted in more correct Type I error rates than MLM-CS and rANOVA when the sphericity assumption was violated and when sample size was $n = 80$ or larger. However, our results differ from Kowalchuk et al.'s (2004) finding that MLM-UN performs similarly to fitting the true covariance structure and, under certain conditions, shows even better Type I error control because MLM-CS had a more correct Type I error rate than MLM-UN when the sphericity assumption was not violated.

There was a conservative bias of rANOVA-GG, especially when the population data was conform to the sphericity assumption, when the number of measurement occasions was large, and when the sample size was small. Therefore, the use of rANOVA-GG cannot be recommended with $n = 20$ and nine measurement occasions when the sphericity assumption is not violated. rANOVA-HF had equal or more correct Type I error rates across all conditions of the simulation study than rANOVA-GG. Thus, the present simulation study supports the use of rANOVA-HF instead of rANOVA-GG.

Further research may consider the investigation of between-group effects and especially the investigation of between-group × within-group interaction effects for up to nine measurement occasions. Moreover, the statistical power of MLM and rANOVA should also be investigated for a large number of measurement occasions. There are, of course, several other issues, as, for example, the combined effect of the number of data levels and measurement occasions on Type I error rates and statistical power. The high flexibility of MLM as a method of multivariate analysis results in a specific responsibility of researchers in the application of this method. However, given a responsible use of MLM, the considerable flexibility is an important advantage of this method.

Two general recommendations follow from the results of the present simulation study with respect to the violation of the sphericity assumption: (1) Use rANOVA-HF, especially when sample sizes are small and when the number of measurement occasions is large. (2) Use MLM-UN when the sample size is at least $n = 80$. Accordingly, when there are reasons for performing MLM-UN instead of rANOVA-HF as, for example, when there is a substantial number of missing values or a more complex multi-level structure of the data, sample sizes of at least $n = 80$ should be investigated. And the most general recommendation from the present results is the following: Whether the sphericity assumption is violated or not, MLM-UN should not be used in combination with sample sizes of about $n = 60$ or smaller and nine or more measurement occasions unless a correction of the substantial progressive bias of this method is available.

**Conflict of Interest**

*The authors declare that the research was conducted in the absence of any commercial or financial relationships that could be construed as a potential conflict of interest.*